\documentclass[]{aa}

\usepackage{txfonts}
\usepackage{graphicx}
\usepackage{hyperref}
\usepackage{xcolor}
\usepackage{algorithm}
\usepackage{algorithmicx}
\usepackage[inline]{enumitem}
\usepackage{natbib}
\bibpunct{(}{)}{;}{a}{}{,}

\hypersetup{colorlinks=true,citecolor=blue,linkcolor=blue}

\begin{document}

\title{
    Growing the seeds of pebble accretion through planetesimal accretion
    }

\author{
    Sebastian~Lorek\inst{1},
    Anders~Johansen\inst{1,2}
    }

\institute{
    Centre for Star and Planet Formation,
    Globe Institute, 
    University of Copenhagen,
    {\O}ster Voldgade 5–7, 
    DK-1350 Copenhagen, 
    Denmark \\
    \email{sebastian.lorek@sund.ku.dk}
    \and
    Lund Observatory, 
    Department of Astronomy and Theoretical Physics, 
    Lund University,
    Box 43, 
    221 00 Lund,
    Sweden
    }

\date{Received ; accepted }

\abstract{
We explore the growth of planetary embryos by planetesimal accretion up to and beyond the point where pebble accretion becomes efficient at the so-called Hill-transition mass. Both the transition mass and the characteristic mass of planetesimals formed by the streaming instability increase with increasing distance from the star. We developed a model for the growth of a large planetesimal (embryo) embedded in a population of smaller planetesimals formed in a filament by the streaming instability. The model includes in a self-consistent way the collisional mass growth of the embryo, the fragmentation of the planetesimals, the velocity evolution of all involved bodies, as well as the viscous spreading of the filament. We find that the embryo accretes all available material in the filament during the lifetime of the protoplanetary disc only in the inner regions of the disc. In contrast, we find little or no growth in the outer parts of the disc beyond 5--10 AU. Overall, our results demonstrate very long timescales for collisional growth of planetesimals in the regions of the protoplanetary disc where giant planets form. As such, in order to form giant planets in cold orbits, pebble accretion must act directly on the largest bodies present in the initial mass-function of planetesimals with little or no help from mutual collisions.
}

\keywords{
    Methods: numerical --
    Planets and satellites: formation
    }

\maketitle


\section{Introduction}
\label{sec:introduction}

The classic picture for the formation of planets requires growth over several orders of magnitude in size. Starting from micrometre-sized dust and ice grains, coagulation produces millimetre-sized pebbles. Collisional and dynamical processes, such as bouncing, fragmentation, and radial drift, limit the maximum particle size \citep{Blum2008,Guettler2010,Zsom2010,Krijt2015} and the formation of larger bodies is effectively prevented. Porosity in combination with an increased stickiness of ice could bypass these barriers \citep{Wada2009,Okuzumi2012,Kataoka2013}. However, it requires that ice is indeed stickier than rocky material \citep{Gundlach2015,Arakawa2021,Schraepler2022}, which might not necessarily be the case \citep{Musiolik2019,Kimura2020}, and that the initial grains are sub-micron in size.

An alternative mechanism that has been extensively studied since its discovery invokes the concentration of pebbles through streaming instability and the subsequent gravitational collapse of dense filament-like structures that converts ${\sim}\mathrm{mm}$-sized pebbles directly to ${\sim}\mathrm{km}$-sized planetesimals \citep[e.g.][]{Youdin2005,Johansen2007,Johansen2014,Simon2016,Schaefer2017,Abod2019}. These planetesimals then would grow to planet-sized bodies through runaway and oligarchic growth.

Runaway growth occurs when a planetesimal that is slightly more massive than the rest of the bodies accretes more efficiently through gravitational focusing. The mass ratio between two bodies then increases with time and the more massive one quickly outgrows the other planetesimals. As the growing body becomes more massive, it starts to gravitationally stir the surrounding planetesimals which reduces gravitational focusing and runaway growth ceases \citep{Ida1993,Kokubo1996,Ormel2010a}. Eventually, a number of planetary embryos form which grow in an oligarchic fashion by accreting the planetesimals in their respective feeding zones until reaching their isolation mass \citep{Kokubo1998,Kokubo2000}. In the final assembly of planets, these bodies grow by collisions and when reaching a threshold mass of ${\sim}10\,M_\oplus$ they start to accrete gas from the surrounding nebula to form the terrestrial and giant planets.

Planetesimal accretion has long thought to be the main pathway of planet formation. However, accretion is only efficient if planetesimals are small. For small planetesimals, of the order of a few kilometres in size at most, gas drag damps eccentricities and inclinations, boosting the accretion rate. However, it is uncertain if planetesimals actually formed small or if they were large to begin with, typically around $100\,\mathrm{km}$ in diameter \citep{Morbidelli2009,Weidenschilling2011,Johansen2015}. Evidence for the latter case is seen not only in the size distribution of the asteroid belt \citep{Bottke2005} and the cold classical Kuiper belt objects \citep{Kavelaars2021}, which are most likely the unaltered remnants of the planetesimals that formed in the outer Solar System, but also in the absence of large craters on Pluto, which indicates a lack of bodies ${\lesssim}1$ to $2\,\mathrm{km}$ in diameter \citep{Singer2019}. Furthermore, numerical studies of planetesimal formation through the streaming instability point towards a large initial size \citep{Youdin2005,Johansen2007}. The fragmentation of dense pebble filaments into planetesimals results in an initial mass-function (IMF) of planetesimals that is well described by a power-law with exponential cut-off for bodies exceeding a characteristic mass. The characteristic mass translates to a characteristic size of ${\sim}100\,\mathrm{km}$ at a heliocentric distance of the asteroid belt and the largest bodies that form through this process are roughly the size of Ceres (${\sim}10^{-4}\,M_\oplus$) \citep{Simon2016,Schaefer2017,Abod2019,Li2019}.

Because of the long timescales for planetesimal accretion, an efficient formation of terrestrial planets and the cores of giant planets within the lifetime of the protoplanetary disc of typically only a few $\mathrm{Myr}$ \citep{Haisch2001} is problematic. \citet{Johansen2019b} explored the conditions for forming the cores of the giant planets through planetesimal accretion. Their model focuses on the growth track of a single migrating protoplanet sweeping through a population of planetesimals. They found that their fiducial model with constraints from the Solar System of
\begin{enumerate*}[label=(\roman*)]
\item a primordial population of planetesimals of a few hundred Earth masses,
\item a characteristic planetesimal size of ${\sim}100\,\mathrm{km}$, and
\item a weakly turbulent protoplanetary disc
\end{enumerate*}
allows protoplanets to grow to only ${\sim}0.1\,M_\oplus$ within the disc lifetime of $3\,\mathrm{Myr}$. Allowing for a massive disc of planetesimals of ${\sim}1000\,M_\oplus$ produces close-in giant planets, but fails to form cold giant planets, such as Jupiter or Saturn in the Solar System, unless the planetesimal size and turbulence strength are reduced at the same time. Their conclusion is that unless ignoring all three constraints, planetesimal accretion is insufficient to grow the cores of giant planets.

In contrast, fast growth can be achieved by the accretion of pebbles that are ubiquitous in the protoplanetary disc. Processes like the streaming instability that explain planetesimal formation would convert between ${\sim}10\,\%$ and up to $80\,\%$ of the pebble mass trapped in filaments to planetesimals \citep{Abod2019}. The remnant pebbles as well as newly forming pebbles in the outer disc, where growth timescales are longer, would then provide a mass reservoir for further growth through pebble accretion. Pebble accretion becomes efficient for sufficiently large embryos above the so-called transition mass \citep{Lambrechts2012}, at which the growth mode changes from slow Bondi accretion to the fast Hill accretion. The transition mass evaluates to ${\sim}2{\times}10^{-3}\,M_\oplus$ at $1\,\mathrm{AU}$ and ${\sim}6{\times}10^{-3}\,M_\oplus$ at $10\,\mathrm{AU}$ \citep{Ormel2010,Lambrechts2012}. Such a body then accretes a large amount of pebbles within a short timescale and can form the terrestrial planets and the cores of the giant planets, consistent with the lifetime of protoplanetary discs \citep{Lambrechts2012,Lambrechts2014,Johansen2017,Johansen2021}. However, a body of ${\sim}10^{-3}$ to $10^{-2}\,M_\oplus$ needs to form in the first place and planetesimal accretion could be the process for that.

In this paper, we investigate if and under which conditions planetesimal accretion would lead to the formation of pebble-accreting embryos. Most simplified models developed to describe the growth of planets start with a narrow annulus of uniformly distributed planetesimals and an embryo in the centre. The growth of the embryo is followed until all planetesimals from the feeding zone are accreted \citep[e.g.][]{Thommes2003,Chambers2006a,Fortier2013}. It is commonly assumed that there is only one planetesimal size, that the embryo mass follows from the transition from runaway to oligarchic growth, and that the feeding zone of the embryo is always populated with planetesimals. In our model, we want to deviate in some aspects from this approach. While we also model the growth of an embryo embedded in a population of planetesimals, we employ a different initial situation where we
\begin{enumerate*}[label=(\roman*)]
\item limit the available mass at a certain location by assuming that it is given by the mass budget of a streaming instability filament and
\item use the streaming instability IMF to derive the characteristic planetesimal size and the size of the embryo.
\end{enumerate*}
We furthermore deviate from the approach of \citet{Johansen2019b}, by
\begin{enumerate*}[label=(\roman*)]
\item ignoring migration,
\item including fragmentation,
\item having a self-consistent treatment of the growth rates and the eccentricity and inclination evolution of embryos, planetesimals, and fragments.
\end{enumerate*}
This way, we study here the growth from planetesimals to planetary embryos, while \citet{Johansen2019b} focused on the later growth stages where migration is important.

\citet{Liu2019} investigates the growth from planetesimals to embryos and beyond through planetesimal and pebble accretion at the water snowline by means of $N$-body simulations. In their work, they test different initial conditions for the planetesimal population of
\begin{enumerate*}[label=(\roman*)]
\item a mono-dispersed population of planetesimals,
\item a poly-dispersed population with IMF from streaming instability simulations, and
\item a two-component population emerging from runaway growth of planetesimals.
\end{enumerate*}
They find that a mono-dispersed population of planetesimals of size $400\,\mathrm{km}$ fails to form planets because growth timescales are too long due to the rapid excitation of eccentricities and inclinations of the planetesimals. In the other two cases, however, the largest body that forms, either because of the IMF or as a result of runaway growth of $100\,\mathrm{km}$-sized planetesimals, grows to several Earth masses firstly by planetesimal accretion and later through pebble accretion when the embryo reaches a mass of ${\sim}10^{-3}$ to $10^{-2}\,M_\oplus$. Our work is complementary to their study because we explore the planetesimal accretion phase at various locations, whereas \citet{Liu2019} focused on a single site, the snowline at $2.7\,\mathrm{AU}$.

The paper outline is as follows. In Sect.~\ref{sec:methods}, we give an outline of our model and our assumptions. In Sect.~\ref{sec:results}, we present the results of planetesimal accretion around a solar-like star, which is our fiducial model. In Sect.~\ref{sec:discussion}, we explore and discuss parameter variations of the fiducial model. Finally, in Sect.~\ref{sec:conclusions}, we summarise and conclude the study.


\section{Methods}
\label{sec:methods}

\subsection{Basic outline}

We use a semi-analytic model to follow the growth of a planetary embryo at a fixed distance $r$ from the central star \citep[e.g.][]{Chambers2006a}. Three different types of bodies are considered. These are
\begin{enumerate*}[label=(\roman*)]
\item an embryo, 
\item a population of planetesimals, and 
\item a population of fragments. 
\end{enumerate*}

The embryo is treated as a single body with mass $M_\mathrm{em}$, radius $R_\mathrm{em}$, eccentricity $e_\mathrm{em}$ and inclination $i_\mathrm{em}$, and surface density $\Sigma_\mathrm{em}$. For the surface density of the embryo, we simply assume that the mass of the embryo is distributed uniformly in an annulus of area $A_\mathrm{em}$ centred at $r$ with a width of $b r_\mathrm{h}$,
\begin{equation}
\Sigma_\mathrm{em}=\frac{M_\mathrm{em}}{2\pi r b r_\mathrm{h}}=\frac{(3M_\odot)^{1/3}}{2\pi r^2 b}M_\mathrm{em}^{2/3},
\label{eq:embryosurfacedensity}
\end{equation}
where $r_\mathrm{h}{=}r\left(M_\mathrm{em}/(3M_\odot)\right)^{1/3}$ is the Hill radius \citep{Chambers2006a}. The value $b{=}10$ corresponds to the typical spacing of isolated embryos which has been shown in $N$-body simulations to be ${\sim}10$ Hill radii \citep{Kokubo1998}.

A single planetesimal has mass $m_\mathrm{p}$, radius $R_\mathrm{p}$, and the population has a surface density $\Sigma_\mathrm{p}$. We take the root-mean-square eccentricity and inclination to describe the orbits of the planetesimals. The fragments are treated in the same way as the planetesimals with a mass $m_\mathrm{fr}$ and radius $R_\mathrm{fr}$ for a single fragment, and surface density $\Sigma_\mathrm{fr}$, root-mean-square eccentricity and inclination for the population. 

\subsection{Mass and surface density evolution, fragmentation}

The embryo grows by accreting planetesimals and fragments. The accretion rate of the embryo can be written as
\begin{equation}
\dot{M}_\mathrm{em}^{(k)}=h_{\mathrm{em},k}^2r^2\Omega_\mathrm{K}\Sigma_{k}P_\mathrm{col},
\end{equation}
where $k$ stands for either planetesimals or fragments. Furthermore, we have the reduced mutual Hill radius $h_{\mathrm{em},k}{=}\left(\left(M_\mathrm{em}+m_k\right)/\left(3M_\odot\right)\right)^{1/3}$, the Keplerian frequency $\Omega_\mathrm{K}$, and a dimensionless collision rate $P_\mathrm{col}$ that is a function of the sizes of the colliding bodies, and their mutual eccentricities and inclinations \citep{Inaba2001,Chambers2006a}.

We are now able to formulate the evolution equations for the surface densities. For the embryo surface density, we differentiate Eq.~\ref{eq:embryosurfacedensity} with respect to time and get
\begin{equation}
\dot{\Sigma}_\mathrm{em}^{(k)}=\frac{(3M_\odot)^{1/3}}{3\pi b r^2 M_\mathrm{em}^{1/3}}\dot{M}_\mathrm{em}^{(k)}=\frac{2}{3}\frac{\Sigma_\mathrm{em}}{M_\mathrm{em}}\dot{M}_\mathrm{em}^{(k)},
\label{eq:surfacedensityem}
\end{equation}
 for the change of surface density due to the accretion of bodies from population $k$.

To derive the evolution of planetesimal and fragment surface densities, we first need to derive the amount of fragments produced in a collision between planetesimals. To do so, we first calculate the number of collisions per unit time between planetesimals
\begin{equation}
\dot{N}_\mathrm{p}=h_{\mathrm{p},\mathrm{p}}^2r^2\Omega_\mathrm{K}N_{\mathrm{s},\mathrm{p}}^2A_\mathrm{p}P_\mathrm{col},
\label{eq:collisionratep}
\end{equation}
where $N_{\mathrm{s},\mathrm{p}}{=}\Sigma_\mathrm{p}/m_\mathrm{p}$ is the surface number density of planetesimals and $P_\mathrm{col}$ is the collision rate between planetesimals \citep{Inaba2001}. Initially, there are no fragments. When planetesimals are excited to high enough eccentricity, such that $ev_\mathrm{K}{\approx}v_\mathrm{esc}$, collisions become disruptive and fragments are produced. Typically this results in a collisional cascade with a size distribution of fragments, however, here we use a typical fragment size of $0.5\,\mathrm{km}$ to represent the fragment population. The fragmentation model of \citet{Kobayashi2010} allows us to determine the total mass of fragments $\Delta M_\mathrm{fr}$ that is produced in a disruptive collision between two planetesimals. The value of $\Delta M_\mathrm{fr}$ depends on the ratio of impact energy and material-dependent critical disruption energy of the planetesimals. By multiplying the collision rate of planetesimals with $\Delta M_\mathrm{fr}$ we get the mass production rate of fragments
\begin{equation}
\dot{M}_\mathrm{fr}^{+}=\dot{N}_\mathrm{p}\Delta M_\mathrm{fr}.
\label{eq:productionratefr}
\end{equation}

Our approach is to keep the total mass of solids constant, that is we have the condition
\begin{equation}
M_\mathrm{em}+M_\mathrm{p}+M_\mathrm{fr}=\mathrm{const.},
\label{eq:totalmass}
\end{equation}
where the total mass of planetesimals $M_\mathrm{p}$ is given by $\Sigma_\mathrm{p}A_\mathrm{p}$, where $A_\mathrm{p}{=}2\pi r\Delta r$ is the area of the annulus of width $\Delta r$ that the planetesimals occupy; and likewise for the fragments. The initial width of the annulus is $\eta r$ (see below) for planetesimals and fragments, but the annuli widen diffusively with time because of excitation of eccentricities and inclinations due to viscous stirring \citep{Ohtsuki2003,Tanaka2003}.

With the assumption of constant total mass, we can now formulate the evolution equations for the surface densities given the mass accretion rate of the embryo $\dot{M}_\mathrm{em}^{(k)}$ and the production rate of fragments by taking the time derivative of Eq.~\ref{eq:totalmass}
which gives
\begin{equation}
\dot{M}_\mathrm{em}+\dot{M}_\mathrm{p}+\dot{M}_\mathrm{fr}=0.
\end{equation}
Substituting the total mass changes with surface densities and areas, we get that the surface density of planetesimals reduces due to accretion of planetesimals by the embryo and due to fragmentation
\begin{equation}
\dot{\Sigma}_\mathrm{p}=-\frac{3A_\mathrm{em}}{2A_\mathrm{p}}\dot{\Sigma}_\mathrm{em}^{(\mathrm{p})}-\frac{\dot{M}_\mathrm{fr}^{+}}{A_\mathrm{p}}.
\label{eq:surfacedensityp}
\end{equation}
The area ratio appears because of the mass conservation condition. Likewise, the surface density of fragments evolves according to
\begin{equation}
\dot{\Sigma}_\mathrm{fr}=-\frac{3A_\mathrm{em}}{2A_\mathrm{fr}}\dot{\Sigma}_\mathrm{em}^{(\mathrm{fr})}+\frac{\dot{M}_\mathrm{fr}^{+}}{A_\mathrm{fr}}.
\label{eq:surfacedensityfr}
\end{equation}
The set of Eqs.~\ref{eq:surfacedensityem}, \ref{eq:collisionratep}, \ref{eq:productionratefr}, \ref{eq:surfacedensityp}, and \ref{eq:surfacedensityfr} fully describe the mass growth of the embryo and the conversion of planetesimals into fragments.

We emphasise that the embryo does not grow to the isolation mass in the classical sense by accreting all the material in the expanding feeding zone. Instead, growth is limited by the available mass contained in the annulus of width $\Delta r$. We consider this approach suited for studying the growth of an embryo in an isolated filament formed through streaming instability where a fixed amount of mass is converted to planetesimals in a confined narrow ring. Furthermore, while the embryo grows by the accretion of planetesimals and fragments, the mass distribution of planetesimals and fragments does not evolve over time owing to our choice of representing those populations by bodies of a characteristic mass. 

\subsection{Velocity evolution}

The velocity distributions, that is the eccentricities and inclinations, of the bodies evolve through viscous stirring and dynamical friction. To take this into account, we use the rate equations from \citet{Ohtsuki2002} for the root-mean-square eccentricities and inclinations. We include viscous stirring and dynamical friction between all populations with the exception that the single embryo is not interacting with itself. Gas drag dampens the orbits of the bodies and we include damping for the embryo, the planetesimals, and the fragments \citep{Adachi1976,Inaba2001}. We do not include turbulent stirring of planetesimals through the disc gas.

\subsection{Protoplanetary disc}

We use the self-similar solution for a viscously evolving $\alpha$-disc \citep{Shakura1973,LyndenBell1974}. The disc is heated by stellar irradiation with a temperature profile of
\begin{equation}
T=150\,\mathrm{K}\left(\frac{M_\star}{M_\odot}\right)^{-1/7}\left(\frac{L_\star}{L_\odot}\right)^{2/7}\left(\frac{r}{\mathrm{AU}}\right)^{-3/7}
\end{equation}
\citep{Ida2016}. The viscously heated part of the disc would initially extend to ${\sim}5\,\mathrm{AU}$ for our fiducial parameter choices \citep{Ida2016}. However, we neglect viscous heating here
\begin{enumerate*}[label=(\roman*)]
\item because recent work indicates that irradiation rather than viscous heating might be the relevant heat source for protoplanetary discs \citep{Mori2019,Mori2021}, and
\item because we verified by running a model with a viscous temperature profile where it applies that the choice of temperature profile has negligible impact on the planetesimal accretion process studied here. The filament mass (see Sect.~\ref{sec:filamentmass}) and the initial radii of planetesimals and embryos (see Sect.~\ref{sec:initialmassesofplanetesimalsandembryo}) vary only within a factor of unity which does not affect the results.
\end{enumerate*}

The viscosity in the $\alpha$-disc model is $\nu{=}\alpha c_\mathrm{s}^2\Omega_\mathrm{K}^{-1}$, where we use $\alpha{=}10^{-2}$. The value of $\alpha$ determines the viscous evolution timescale of the disc and the accretion of the gas onto the star. The value is consistent with what is determined from observations of protoplanetary discs \citep{Hartmann1998}. For the temperature profile used here, the viscosity is a power-law in radial distance, $\nu{\propto}r^{\gamma}$, with an exponent of $\gamma{=}15/14$. 

The surface density profile of the self-similar solution for a power-law viscosity is
\begin{equation}
\Sigma_\mathrm{gas}=\frac{\dot{M}_{\star,0}}{3\pi\nu_1}\left(\frac{r}{r_1}\right)^{-\gamma}\tau^{(5/2-\gamma)/(2-\gamma)}\exp{\left[-\frac{1}{\tau}\left(\frac{r}{r_1}\right)^{2-\gamma}\right]}.
\end{equation}
where $t_\mathrm{vis}=r_1^2/\left(3(2-\gamma)^2\nu_1\right)$ is the characteristic time for the viscous evolution, $\tau{=}\left(t/t_\mathrm{vis}\right)+1$, $r_1$ is the characteristic radius of the disc, and $\nu_1{=}\nu(r_1)$ is the viscosity at distance $r_1$. For the initial mass accretion rate, we use $\dot{M}_{\star,0}{=}10^{-7}\,M_\odot\,\mathrm{yr}^{-1}$ at $0.5$ Myr, which corresponds to a typical class-II object \citep{Hartmann1998,Hartmann2016}. The total disc mass is set to be $10\,\%$ of the stellar mass and by integrating the surface density at $t{=}0$ from the inner edge of the disc, which we set to $0.1\,\mathrm{AU}$, to infinity, we can determine the characteristic disc radius, which is ${\sim}72$ AU in our fiducial case.

\subsection{Formation time of planetesimals and embryo}

Our model start at time $t_0$ when planetesimals are expected to have formed. To estimate $t_0$, we assume that the planetesimal accretion phase takes place in the class-II phase of the disc, ${\sim}0.5\,\mathrm{Myr}$ after star formation \citep{Evans2009,Williams2011}. 

To form planetesimals, dust first grows by coagulation to pebbles. The growth timescales $t_\mathrm{grow}{=}R/\dot{R}$ of the dust is
\begin{equation}
t_\mathrm{grow}=\frac{2}{\sqrt{\pi}\epsilon_\mathrm{g} Z \Omega_\mathrm{K}},
\label{eq:growthtimescale}
\end{equation}
where $\epsilon_\mathrm{g}{\approx}0.5$ is a coagulation efficiency and $Z$ is the solid-to-gas ratio of the dust in the disc \citep{Birnstiel2012,Lambrechts2014}. The time for grains of radius $R_0$ to grow to pebbles of radius $R_\mathrm{peb}$ is found to be
\begin{equation}
\Delta t\approx t_\mathrm{grow}\ln\left(\frac{R_\mathrm{peb}}{R_0}\right)
\label{eq:pebblegrowthtime}
\end{equation}
\citep{Lambrechts2014}. Dust growth is limited by radial drift and fragmentation \citep{Birnstiel2012}, and we use the minimum of both for the pebbles. The fragmentation-limited Stokes number is
\begin{equation}
\tau_{\mathrm{s},\mathrm{frag}}=\frac{v_\mathrm{frag}^2}{2\alpha_\mathrm{t}c_\mathrm{s}^2},
\end{equation}
where we set the collision velocity of similar-sized pebbles driven by turbulence \citep{Ormel2007} equal to the fragmentation threshold velocity $v_\mathrm{frag}$. The value of $v_\mathrm{frag}$ ranges from ${\sim}1\,\mathrm{m}\,\mathrm{s}^{-1}$ for silicate pebbles \citep{Blum2008} to ${\sim}10\,\mathrm{m}\,\mathrm{s}^{-1}$ for icy pebbles \citep{Gundlach2015}. However, more recent studies have shown that ice might not be as sticky as previously thought \citep[e.g.][]{Musiolik2019,Kimura2020} and that the tensile strength of ice aggregates is comparable to the tensile strength of silicates \citep{Gundlach2018}. We  therefore use a common fragmentation threshold velocity of $v_\mathrm{frag}{=}1\,\mathrm{m}\,\mathrm{s}^{-1}$  for fragmentation-limited growth in our study. The turbulent collision velocity depends on the midplane turbulence $\alpha_\mathrm{t}$, which is different from the gas $\alpha$ that drives the viscous evolution of the protoplanetary disc. The value of $\alpha_\mathrm{t}$ obtained from observations of the dust component of protoplanetary discs ranges from ${\sim}10^{-5}$ to a few times $10^{-4}$ \citep{Pinte2016,Villenave2022}. We decided to use a value of $\alpha_\mathrm{t}{=}10^{-4}$ in our study.

In the drift-limited case, the Stokes number is
\begin{equation}
\tau_{\mathrm{s},\mathrm{drift}}=\frac{3\sqrt{\pi}}{4}\frac{\epsilon_\mathrm{g}Z}{\eta},
\end{equation}
which is obtained from setting the growth timescale of the pebbles (Eq.~\ref{eq:growthtimescale}) equal to the drift timescale $t_\mathrm{drift}{=}r/v_r$ with $v_r{=}2\eta v_\mathrm{K}\tau_\mathrm{s}$ being the radial velocity of the pebbles.

The pebbles are subsequently concentrated by the streaming instability and form planetesimals through the gravitational collapse of dense particle filaments. Depending on the size of the pebbles that form, the streaming instability takes about ten (for big pebbles, $\tau_\mathrm{s}{\approx}0.3$) to a few thousand (for small pebbles, $\tau_\mathrm{s}{\approx}10^{-3}{-}10^{-2}$) local orbital periods to create dense filaments that subsequently fragment gravitationally into planetesimals \citep{Yang2014,Yang2017,Li2018,Li2019}. We calculate the pebble growth time at distance $r$ according to Eq.~\ref{eq:pebblegrowthtime} and the streaming instability timescale as $t_\mathrm{SI}{\sim}500\,2\pi\Omega_\mathrm{K}^{-1}$ and add both to the $0.5\,\mathrm{Myr}$ to obtain the initial time $t_0$ for our simulations.

\subsection{Filament mass}
\label{sec:filamentmass}

\begin{figure}
\includegraphics[width=\columnwidth]{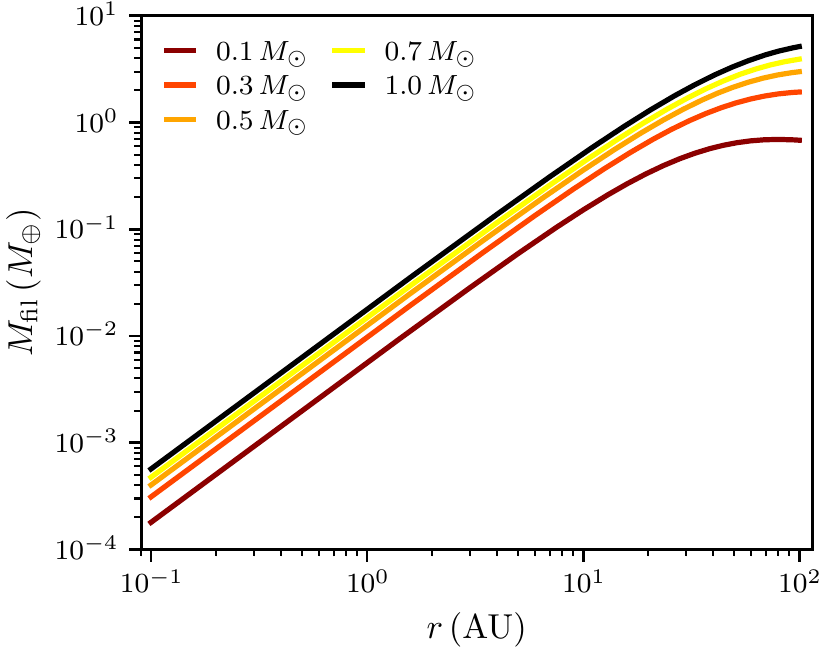}
\caption{Total mass of filaments formed through streaming instability. The mass of the filament $M_\mathrm{fil}{=}2\pi r \Sigma_\mathrm{gas}Z_\mathrm{fil}\eta r$ increases with distances from ${\lesssim}10^{-3}\,M_\oplus$ in the inner disc to almost $10\,M_\oplus$ in the outer disc. $M_\mathrm{fil}$ is the maximum mass the embryo can obtain by accreting planetesimals. Colour coded are the filament masses for different stellar masses.}
\label{fig:filament}
\end{figure}

The initial conditions of the planetesimal population are derived in the framework of planetesimal formation through the streaming instability. We assume that the streaming instability forms a dense filament of pebbles which fragments into planetesimals. The typical radial width of a filament is ${\sim}\eta r$, where $\eta$ is related to the pressure gradient of the disc gas and $r$ is the distance from the star \citep{Yang2014,Liu2019,Gerbig2020}. This length scale can be thought of as the length scale over which the Keplerian flow adjusts to the gas flow. We set the solid-to-gas ratio in the filament to $Z_\mathrm{fil}{=}0.1$ and the mass contained in one filament is therefore $M_\mathrm{fil}{=}2\pi r \Sigma_\mathrm{gas} Z_\mathrm{fil} \eta r$ \citep{Liu2019}. Because not all pebbles are converted into planetesimals, we introduce a planetesimal formation efficiency $p_\mathrm{eff}$. The total mass of planetesimals is then $p_\mathrm{eff}\,M_\mathrm{fil}$ \citep{Liu2019}. For an optimistic upper limit on embryo growth, we assume $p_\mathrm{eff}{=}1$ throughout the study. Figure~\ref{fig:filament} shows the mass of the filaments as a function of distance for different stellar masses.

\subsection{Initial masses of planetesimals and embryo}
\label{sec:initialmassesofplanetesimalsandembryo}

\begin{figure}
\includegraphics[width=\columnwidth]{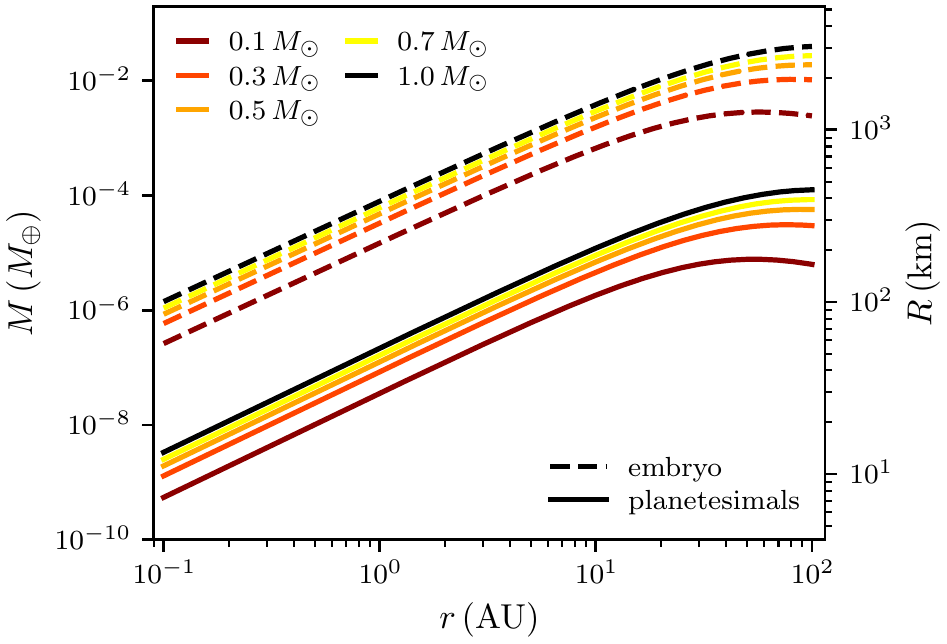}
\caption{Initial masses of planetesimals and embryos. The mass of planetesimals (solid) and embryos (dashed) at initial time $t_0$, which ranges from ${\sim}0.5\,\mathrm{Myr}$ in the inner disc to ${\sim}1.5\,\mathrm{Myr}$ in the outer disc, is shown. The planetesimal size is the characteristic size, which follows from the streaming instability initial mass function, and the embryo size is the single most massive body. Colour coded are the initial sizes for different stellar masses.}
\label{fig:initialsizes}
\end{figure}

The initial sizes of planetesimals and embryos follow from the initial mass-function found in streaming instability simulations. The IMF is a power-law with an exponential cut-off above a characteristic mass,
\begin{equation}
\frac{N_>(m)}{N_\mathrm{tot}}=\left(\frac{m}{m_\mathrm{min}}\right)^{-p}\exp\left[\left(\frac{m_\mathrm{min}}{m_\mathrm{p}}\right)^{q}-\left(\frac{m}{m_\mathrm{p}}\right)^{q}\right]
\label{eq:streaminginstabilityIMF}
\end{equation}
\citep{Schaefer2017}. The slope of the power-law is $p{\approx}0.6$ and the steepness of the cut-off is $q{\approx}0.4$ \citep{Simon2016,Schaefer2017}.  The IMF is top heavy which means that most of the mass is in the large bodies of characteristic mass. We use the characteristic mass above which the IMF drops exponentially as a proxy for the planetesimal mass, which is
\begin{equation}
m_\mathrm{p}\approx5\times10^{-5}\,M_\oplus\,\left(\frac{Z_\mathrm{fil}}{0.02}\right)^{1/2}\left(\frac{\gamma}{\pi^{-1}}\right)^{3/2}\left(\frac{h}{0.05}\right)^3\left(\frac{M_\star}{M_\odot}\right),
\label{eq:characteristicmass}
\end{equation}
where $\gamma{=}4\pi G\rho_\mathrm{g}\Omega_\mathrm{K}^{-2}$ and $h$ is the aspect ratio of the disc \citep{Liu2020}. To determine the mass of the embryo, we calculate the mass of the single most massive body that forms from the IMF. To do so, we set $N_>(m){=}1$ in  Eq.~\ref{eq:streaminginstabilityIMF} and solve for $m$. The value of $N_\mathrm{tot}$ is found by noting that the total number of bodies will be determined by the smallest bodies. We set the minimum mass to $m_\mathrm{min}=10^{-3}m_\mathrm{p}$ and calculate $N_\mathrm{tot}{=}M_\mathrm{fil}/m_\mathrm{min}$. Figure~\ref{fig:initialsizes} shows the initial mass and size of the embryo and the planetesimals as a function of distance for different stellar masses.

\subsection{Diffusion of planetesimals and fragments}

We include the diffusive widening of the planetesimal and fragment rings due to viscous stirring \citep{Ohtsuki2003,Tanaka2003}. This reduces the surface densities with time which impacts both the accretion and the stirring rates. We assume that the initial width of $\eta r$ increases with time as $\sqrt{2 D t}$, where $D$ is the diffusion coefficient and $t$ is the time, as it is characteristic for a random walk \citep{Liu2019}. The diffusion coefficient is related to the viscous stirring rates of eccentricity and inclination \citep{Ohtsuki2003,Tanaka2003}.


\section{Results}
\label{sec:results}

\begin{table}
\caption{fiducial simulation parameters}
\label{tab:simparameters}
\begin{tabular}{lccl}
\hline
parameter & symbol & value & unit \\
\hline
stellar mass & $M_\star$ & $1$ & $M_\odot$ \\
stellar luminosity & $L_\star$ & $1$ & $L_\odot$ \\
disc mass & $M_\mathrm{disc}$ & $0.1$ & $M_\star$ \\
fragment radius & $R_\mathrm{fr}$ & $0.5$ & $\mathrm{km}$ \\
solid bulk density & $\rho_\bullet$ & 2 & $\mathrm{g}\,\mathrm{cm}^{-3}$ \\
mass accretion rate  ($0.5\,\mathrm{Myr}$) & $\dot{M}_{\star,0}$ & $10^{-7}$ & $M_\odot\,\mathrm{yr}^{-1}$ \\ 
viscous parameter & $\alpha$ & $10^{-2}$ & \\
midplane turbulence & $\alpha_\mathrm{t}$ & $10^{-4}$ & \\
filament solid-to-gas ratio & $Z_\mathrm{fil}$ & $0.1$ & \\
grain size & $R_0$ & $0.1$ & $\mu\mathrm{m}$ \\
\hline
\end{tabular}
\end{table}
In the fiducial model, we use a solar-mass central star with a viscously evolving disc heated by solar irradiation. We place filaments at various distances ranging from $0.1$ to $100\,\mathrm{AU}$ and simulate the growth of the embryo for $10\,\mathrm{Myr}$. Table~\ref{tab:simparameters} summarises the model parameters. We later vary the stellar mass, the solid-to-gas ratio of the filament, and other parameters to explore how they affect the growth of the embryo.

\subsection{Growth of the embryo}

\begin{figure}
\includegraphics[width=\columnwidth]{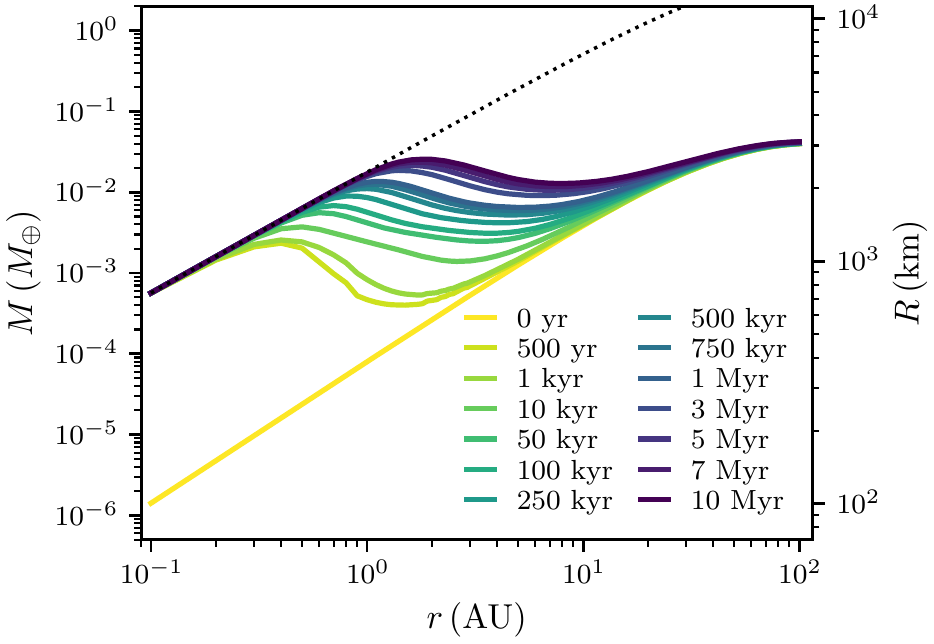}
\caption{Embryo mass at different distances for different time snapshots around a $1\,M_\odot$ star. The embryo mass is shown for different times (relative to $t_0$) as coloured lines. Accretion efficiency decreases with increasing distance and almost ceases outside of ${\sim}20\,\mathrm{AU}$. Overplotted is the maximum mass $M_\mathrm{fil}$ that an embryo can reach (black dotted line).}
\label{fig:massmap}
\end{figure}

Figure~\ref{fig:massmap} shows how the embryo mass evolves with time as a function of distance. The time snapshots are relative to the initial time $t_0$ of the simulation, which ranges from ${\sim}0.5$ Myr in the inner disc to ${\sim}1.5$ Myr in the outer disc. The planetesimal size increases with distance and the initial mass of the embryo is typically a factor $10^{2}$ to $10^{3}$ higher than the planetesimal mass, as shown in Fig.~\ref{fig:initialsizes}. Inside ${\sim}2\,\mathrm{AU}$, the growth timescale is short and embryos reach their final mass by accreting all the available mass in the filament within $10\,\mathrm{Myr}$. Farther out, growth slows down and ceases to almost zero for ${\gtrsim}20\,\mathrm{AU}$.

\begin{figure}
\includegraphics[width=\columnwidth]{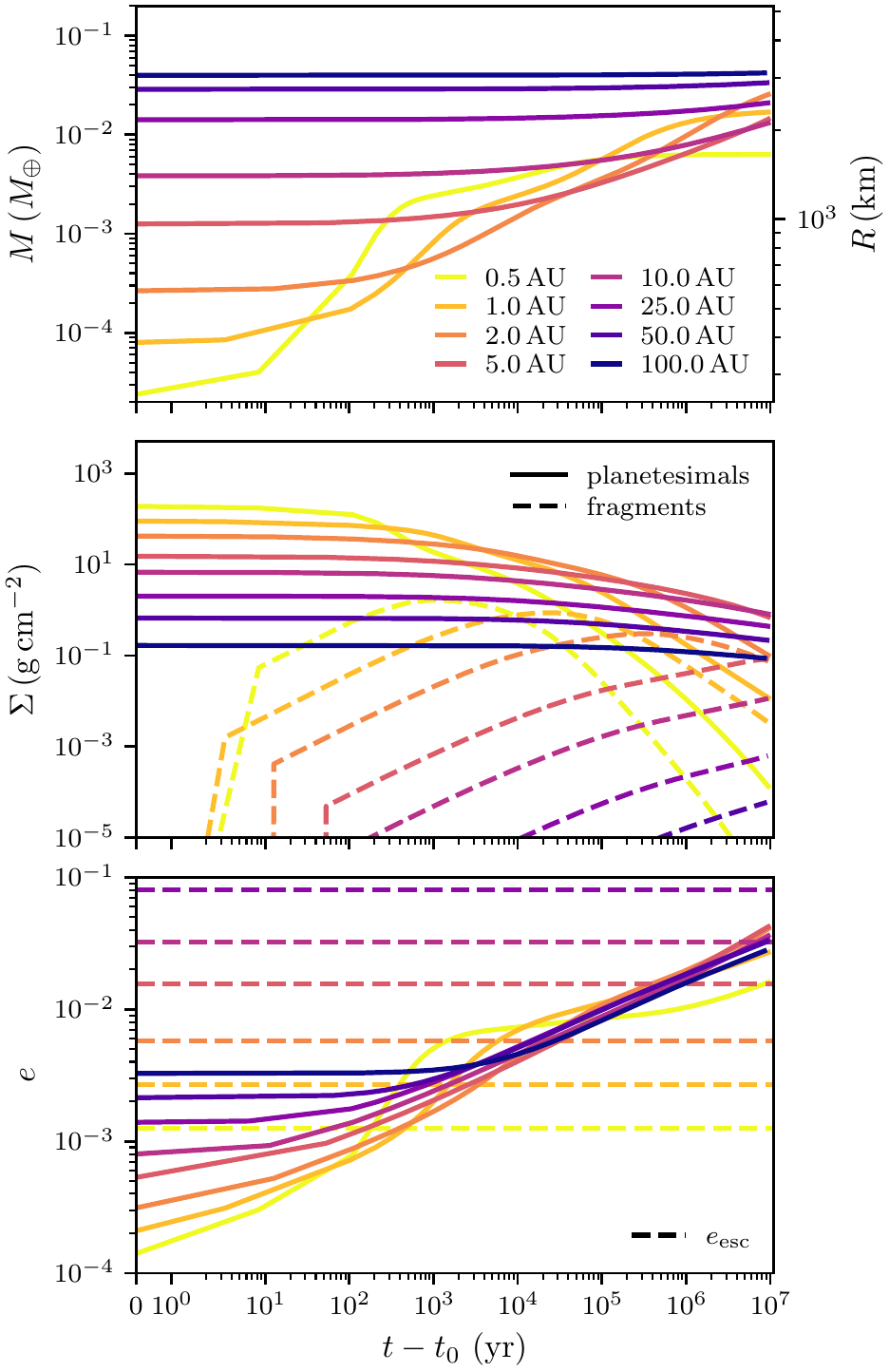}
\caption{Embryo mass, surface densities, and planetesimal eccentricity as function of time at different distances. The different panels are \textit{top}: embryo mass; \textit{middle}: surface densities of planetesimals (solid) and fragments (dashed); \textit{bottom}: planetesimal eccentricity and threshold eccentricity for fragmentation (dashed). Time is expressed relative to the planetesimal formation time $t_0$ for better comparison. The colours correspond to the different distances.}
\label{fig:timemap}
\end{figure}

The reasons for the rapid growth inside ${\sim}2\,\mathrm{AU}$ are the high surface density of planetesimals, which results in a short growth timescale, and the excitation of the eccentricities of the planetesimals, which results in fragmentation and the boost of growth through accretion of fragments. This is visible in Fig.~\ref{fig:timemap}, which shows the time evolution of the surface densities of the planetesimals and the fragments and the eccentricity evolution of the planetesimals in the middle and bottom panels for various distances. The eccentricity at which the collision speed exceeds the escape speed of the planetesimal is approximately given by
\begin{equation}
e_\mathrm{esc}\approx 5\times10^{-3}\left(\frac{\rho_\bullet}{2\,\mathrm{g}\,\mathrm{cm}^{-3}}\right)^{1/2}\left(\frac{R_\mathrm{p}}{100\,\mathrm{km}}\right)\left(\frac{r}{\mathrm{AU}}\right)^{1/2},
\end{equation}
where we set the random speed of the planetesimals ${\sim}ev_\mathrm{K}$ equal to their escape speed. Figure~\ref{fig:timemap} bottom panel shows that within ${\sim}2\,\mathrm{AU}$, the embryo excites the planetesimals above the threshold in short times (${\lesssim}10^{4}\,\mathrm{yr}$). As a consequence, the embryo efficiently accretes the small fragments (which we assume here to have a constant radius of $0.5\,\mathrm{km}$). However, in the outer disc, this effect is negligible because planetesimal eccentricities are not excited enough to result in fragmentation. For example, at $5\,\mathrm{AU}$ and for $200\,\mathrm{km}$ planetesimals (see Fig.~\ref{fig:initialsizes}) this requires eccentricities ${\gtrsim}2{\times}10^{-2}$, which are reached only after ${\sim}1\,\mathrm{Myr}$ (after $t_0$) and later. Therefore, fragmentation, if at all, sets in late and embryo growth is not boosted by fragment accretion as it is the case in the inner disc. At even larger distances, fragmentation plays no role because the stirring of planetesimals by the embryo and by self-stirring is not sufficient to reach $e_\mathrm{esc}$. Therefore, in the outer disc, the long accretion timescale limits the growth.

\subsection{Eccentricity evolution}

\begin{figure}
\includegraphics[width=\columnwidth]{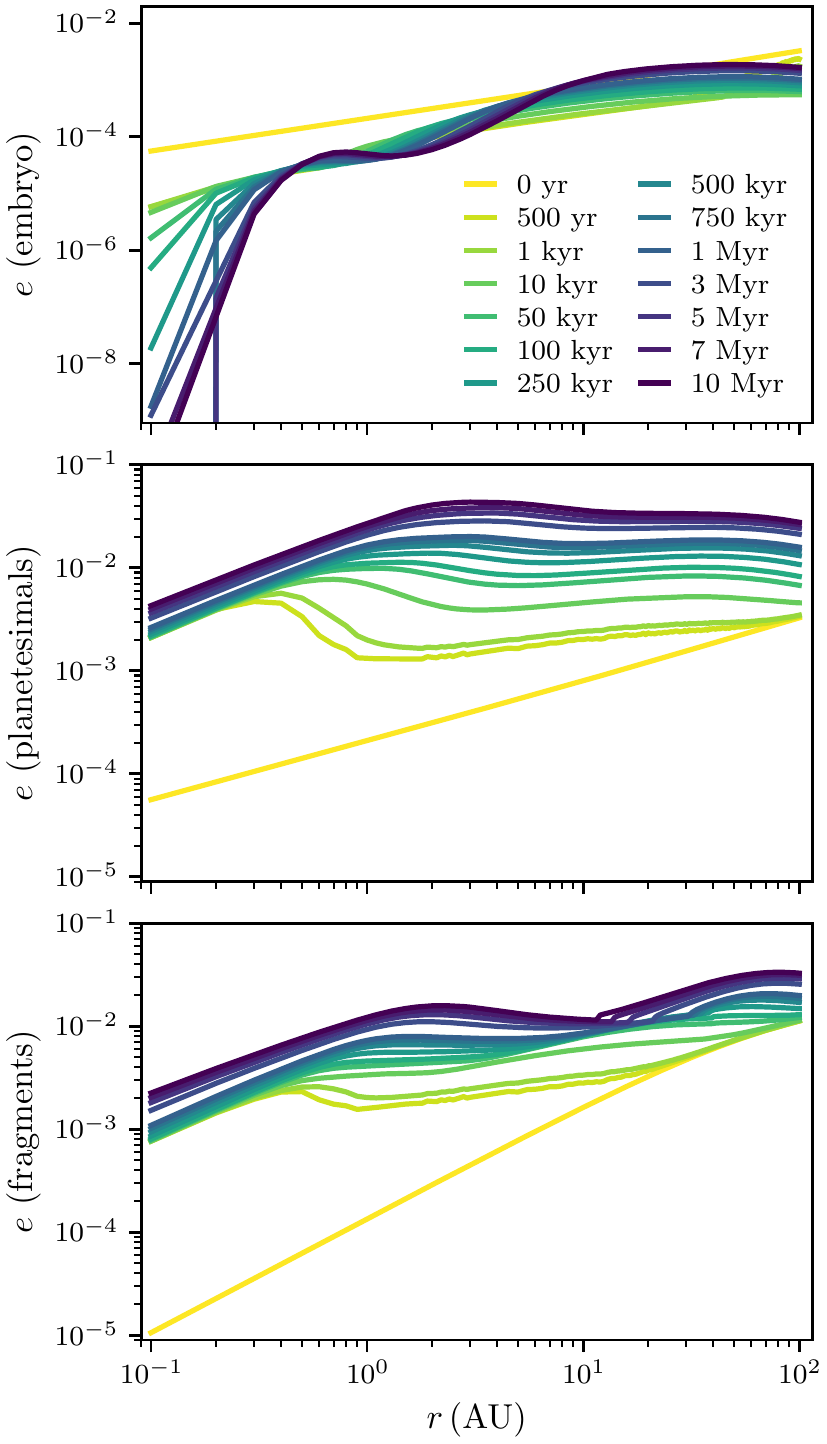}
\caption{Eccentricities of embryos, planetesimals, and fragments at different distances for different times. The different panels show the eccentricities of the embryo (\textit{top}), the planetesimals (\textit{middle}), and the fragments (\textit{bottom}). The eccentricities are shown for different times (relative to $t_0$) as coloured lines. In the inner disc, the bodies follow the equilibrium eccentricity given by the balance of viscous stirring through the embryo and friction with the gas. In the outer disc, eccentricities are determined by viscous stirring with minor damping from gas drag.}
\label{fig:eccentricitymap}
\end{figure}

Figure~\ref{fig:eccentricitymap} shows the eccentricities of embryos, planetesimals, and fragments as a function of distance for different times. Initially, embryos and planetesimals have the same eccentricity and inclination such that $e{=}\eta/2$ and $i/e{=}1/2$, because we assumed that just after formation there has not been enough time for dynamical friction to result in a mass dependent eccentricity. The initial eccentricity of the fragments is set to $10\,\%$ of the escape speed of a planetesimal.

\subsubsection{Planetesimals}

In the inner disc, the eccentricity of the planetesimals is determined by the equilibrium of viscous stirring by the embryo and gas drag, because the gas density is sufficiently high. The damping timescale for gas drag is
\begin{equation}
T_\mathrm{drag}=\frac{1}{e}\frac{2m_\mathrm{p}}{C_\mathrm{D}\pi R_\mathrm{p}^2\rho_\mathrm{gas}r\Omega_\mathrm{K}},
\label{eq:dampingtimescale}
\end{equation}
where $m_\mathrm{p}$ and $R_\mathrm{p}$ are mass and radius of the planetesimal, $\rho_\mathrm{gas}$ is the gas density, and $C_\mathrm{D}$ is the drag coefficient \citep{Adachi1976,Inaba2001}. The timescale on which viscous stirring of planetesimals by an embryo of mass $M_\mathrm{em}$ and surface density $\Sigma_\mathrm{em}$ excites eccentricities is given by
\begin{equation}
T_\mathrm{vis}=\frac{1}{40}\left(\frac{\Omega_\mathrm{K}^2 r^3}{G M_\mathrm{em}}\right)^2 \frac{M_\mathrm{em}e^4}{\Sigma_\mathrm{em}r^2\Omega_\mathrm{K}}
\label{eq:stirringtimescale}
\end{equation}
\citep{Ida1993}. When viscous stirring by the embryo and gas drag on the planetesimal are in equilibrium, the eccentricity can be calculated by setting the damping timescale equal to the stirring timescale,
\begin{equation}
e_\mathrm{eq}=1.7\frac{m_\mathrm{p}^{1/15}M_\mathrm{em}^{1/3}\rho_\bullet^{2/15}}{b^{1/5}C_\mathrm{D}^{1/5}\rho_\mathrm{gas}^{1/5}M_\star^{1/3}r^{1/5}}
\end{equation}
\citep{Thommes2003}. The drag coefficient can be assumed to be $2$ which is valid for planetesimal-sized bodies. The typical spacing of embryos $b$ is of the order of $10$ Hill radii \citep{Kokubo1998} and enters via the embryo surface density $\Sigma_\mathrm{em}{=}M_\mathrm{em}/(2\pi r b r_\mathrm{h})$. From Fig.~\ref{fig:initialsizes}, we can see that the mass of planetesimals and embryos scales with distance as approximately $m_\mathrm{p}{\propto}r^{3/2}$. Because of the distance dependency of $m_\mathrm{p}$ and $\rho_\mathrm{gas}$, $T_\mathrm{drag}{\propto}r^{47/14}$ (for our viscous $\alpha$-disc) increases strongly with distance and the equilibrium eccentricity increases with distance approximately as $e_\mathrm{eq}{\propto}r^{61/70}$, which is the slope of $e$ with $r$ as we see in Fig.~\ref{fig:eccentricitymap} inside of ${\sim}0.8\,\mathrm{AU}$.

In the outer disc, damping by gas drag becomes inefficient because of the lower gas density and because of the large planetesimals. Therefore, viscous stirring by the embryo is the process that determines the planetesimal eccentricity. The viscous stirring of planetesimals by the embryo is expressed as
\begin{equation}
\frac{\mathrm{d}e^2}{\mathrm{d}t}=\frac{e^2}{T_\mathrm{vis}}
\label{eq:stirringrate}
\end{equation}
\citep{Ida1993}. Inserting Eq.~\ref{eq:stirringtimescale}, we can integrate Eq.~\ref{eq:stirringrate} which gives
\begin{equation}
e(t)=e_0\left(1 + \frac{2\left(t-t_0\right)}{T_{{\mathrm{vis},0}}}\right)^{1/4},
\label{eq:ecctime}
\end{equation}
where $T_{\mathrm{vis},0}$ is the viscous stirring timescale for initial planetesimal eccentricity $e_0$. In Fig.~\ref{fig:timemap}, we can see that $e{\propto}t^{1/4}$ for large distances, that is outside of ${\sim}2\,\mathrm{AU}$. Evaluating the scaling with distance, we find that $e{\propto}r^{1/4}$, when taking the $r$ dependency of the embryo mass (initial mass because there is some growth up to ${\sim}20\,\mathrm{AU}$) and of the initial eccentricity (${\propto}r^{4/7}$ because $\eta{\propto}(c_\mathrm{s}/v_\mathrm{K})^2$) into account. However, Fig.~\ref{fig:eccentricitymap} shows a more flat scaling with $r$. The reason is that even though the eccentricity evolution is determined by viscous stirring because the damping timescale is too long to result in equilibrium eccentricities, gas drag still damps the eccentricities.

\subsubsection{Fragments}

The eccentricities of the fragments within ${\sim}1\,\mathrm{AU}$ are given by the equilibrium eccentricity. Because fragments are smaller in size, they are more strongly damped by the gas and hence acquire lower eccentricities. The ratio of fragment size to planetesimal size is ${\sim}10^{-2}$ close to the star, which translates to a ratio of $e_\mathrm{eq}$ to ${\sim}0.4$, as shown in Fig.~\ref{fig:eccentricitymap}. In the outer disc, eccentricities are excited by viscous stirring and damped by gas drag, where equilibrium values might be reached at late times.

\subsubsection{Embryo}

The eccentricity of the embryo is more complex as seen in Fig.~\ref{fig:eccentricitymap} because it is determined through the interplay of viscous stirring through the planetesimals, dynamical friction from planetesimals and fragments, and damping through gas drag. The mass growth further complicates the picture and simple scaling arguments as provided for planetesimals and fragments no longer suffice. However, qualitatively, the embryo keeps a low eccentricity (${\lesssim}10^{-3}$) throughout the simulation. Close to the star, the high gas and planetesimal surface density in combination with the fast growth circularises the orbit. In the outer disc, where no growth occurs, the embryo eccentricity remains close to the initial value experiencing some damping through gas drag and dynamical friction.


\section{Discussion}
\label{sec:discussion}

\subsection{Varying the stellar mass}

The stellar mass affects the mass accretion rate $\dot{M}_\star$, the luminosity $L_\star$, as well as the density and temperature structure of the disc. Here, we investigate the growth of embryos around different stellar masses, ranging from $0.1\,M_\odot$ to $1\,M_\odot$. The mass accretion rate of low-mass stars are also lower. \citet{Manara2012} provide a fit for the mass accretion rate as a function of stellar age and mass. For our chosen initial time of $0.5\,\mathrm{Myr}$, we find $\dot{M}_\star{\propto}M_\star$. \citet{Hartmann2016} find that the mass accretion rate correlates with stellar mass as $\dot{M}_\star{\propto}M_\star^{2.1}$. The linear scaling of \citet{Manara2012}, with $\dot{M}_\star{\propto}M_\star$, results in smaller and more rapidly evolving discs than the quadratic scaling. We run simulations with both relations to scale the initial mass accretion rate of $10^{-7}\,M_\odot\,\mathrm{yr}^{-1}$ for $M_\star{=}1\,M_\odot$ to lower stellar masses. The luminosity scales with mass as $L_\star{\propto}M_\star^{1..2}$ for stellar ages ${\lesssim}10$ Myr \citep{Liu2020} and hence we set the slope of the $L_\star{-}M_\star$-relation to an intermediate value of $1.5$.

\begin{figure}
\includegraphics[width=\columnwidth]{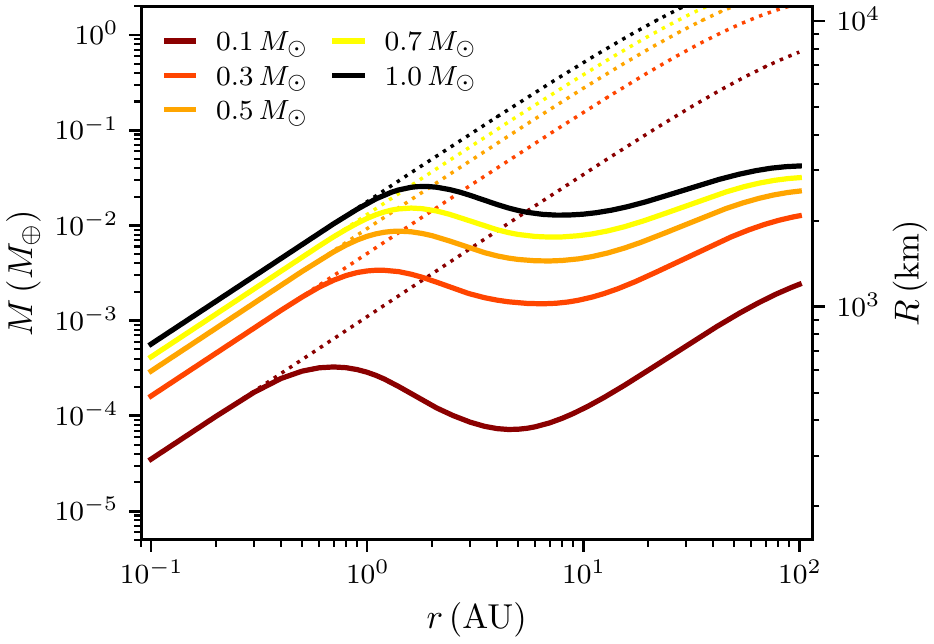}
\caption{Final sizes of embryos for different stellar masses. The embryo mass after $10$ Myr is shown as a function of distances for different stellar masses (colour coded). We used the quadratic scaling $\dot{M}_\star{\propto}M_\star^{2.1}$ \citep{Hartmann2016} to scale the initial mass accretion rate of $10^{-7}\,\mathrm{M_\odot}\,\mathrm{yr}^{-1}$ for lower stellar masses.}
\label{fig:finalsizes}
\end{figure}

\begin{figure}
\includegraphics[width=\columnwidth]{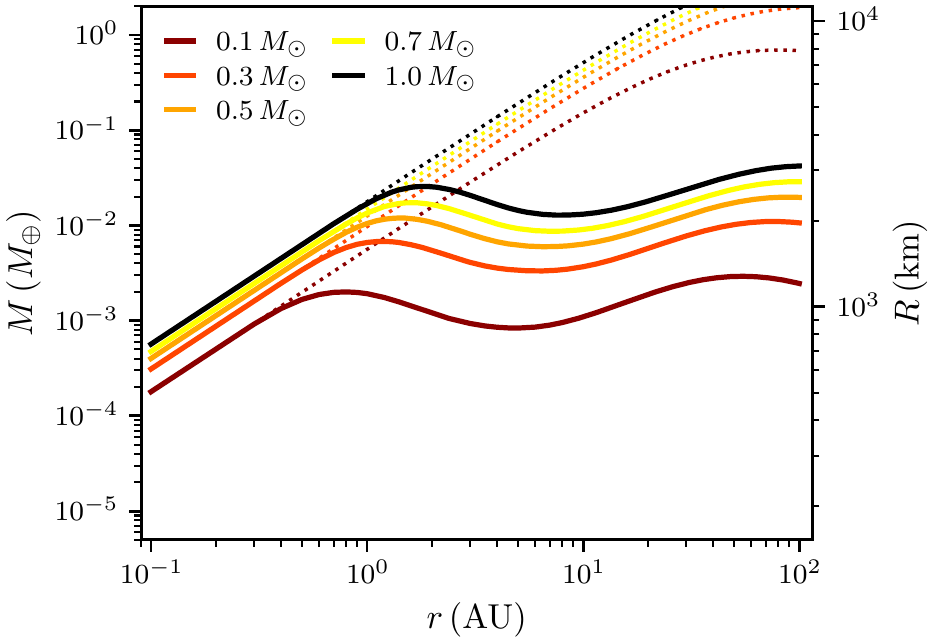}
\caption{Final sizes of embryos for different stellar masses. The embryo mass after $10$ Myr is shown as a function of distances for different stellar masses (colour coded). We used the linear scaling $\dot{M}_\star{\propto}M_\star$ \citep{Manara2012} to scale the initial mass accretion rate of $10^{-7}\,\mathrm{M_\odot}\,\mathrm{yr}^{-1}$ for lower stellar masses.}
\label{fig:finalsizes_linear}
\end{figure}

Figure~\ref{fig:finalsizes} shows the final embryo mass at $10\,\mathrm{Myr}$ as a function of distance for different stellar masses for $\dot{M}_\star{\propto}M_\star^{2.1}$. We find that the maximum distance out to which embryos accrete all available mass of the filament scales with stellar mass, ranging from ${\sim}0.3\,\mathrm{AU}$ for a $M_\star{=}0.1\,M_\odot$ to ${\sim}1\,\mathrm{AU}$ for a $M_\star{=}1\,M_\odot$. Farther out, accretion becomes less efficient and ceases for distances ${\gtrsim}10\,\mathrm{AU}$ for $M_\star{=}0.1\,M_\odot$ and ${\gtrsim}30\,\mathrm{AU}$ for $M_\star{=}1\,M_\odot$. In comparison to the quadratic scaling, the final embryo masses for $\dot{M}_\star{\propto}M_\star$ are shown in Fig.~\ref{fig:finalsizes_linear}. The general finding is the same as for Fig.~\ref{fig:finalsizes}, however, the embryos are more massive for all stellar masses ${<}1\,M_\odot$. The reason for this is that $\Sigma_\mathrm{gas}{\propto}\dot{M}_\star$. For the shallower scaling, the discs around the lower mass stars have higher surface densities and hence the masses of the filaments are higher, which consequently leads to higher masses of embryos and more mass available for the embryos to accrete.

\subsection{Varying the filament solid-to-gas ratio}

\begin{figure}
\includegraphics[width=\columnwidth]{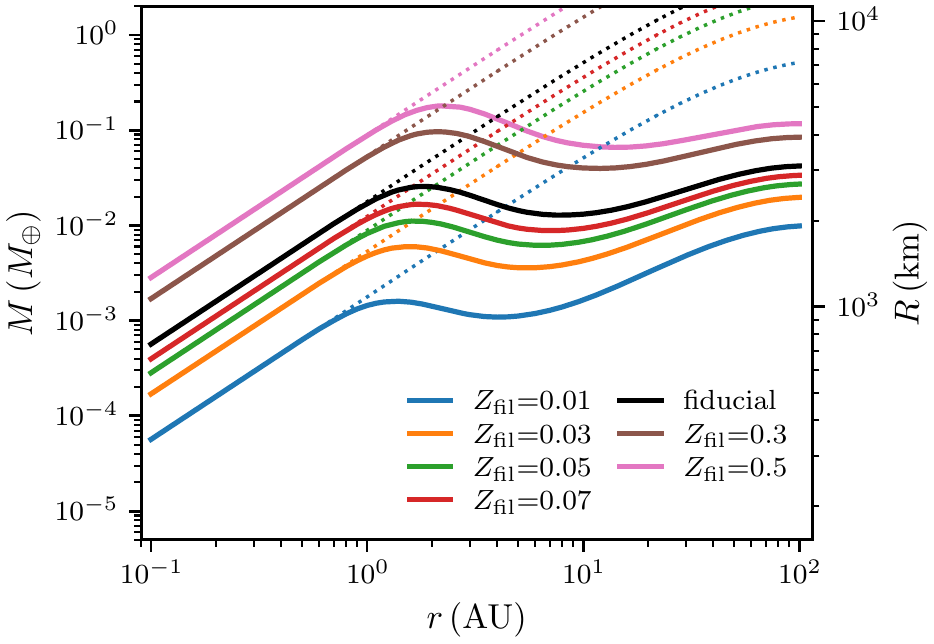}
\caption{Final sizes of embryos for different filament solid-to-gas ratios. The embryo mass after $10$ Myr is shown as a function of distances for a $1\,M_\odot$ star for different values of $Z_\mathrm{fil}$ (coloured lines). The fiducial case (solid black line) as well as the filament mass (dotted lines) are shown for comparison.}
\label{fig:finalsizes_Zfil}
\end{figure}

Figure~\ref{fig:finalsizes_Zfil} shows the outcome of our model for different values of the filament solid-to-gas ratio. This $Z_\mathrm{fil}$ provides a minimum value for how much pebble mass is turned into planetesimals (assuming that the planetesimal formation efficiency is $100\,\%$, which we use in our model). We vary $Z_\mathrm{fil}$ from the canonical value of $Z_\mathrm{fil}{=}0.01$ to $Z_\mathrm{fil}{=}0.5$, which means that the mass in planetesimals would be $50\,\%$ of gas mass at distance $r$. We find that varying the value of $Z_\mathrm{fil}$ does not change the general picture of efficient growth for distances ${\lesssim}1$ to $2\,\mathrm{AU}$. The final embryo masses vary according to the value of $Z_\mathrm{fil}$ simply because the filaments are more massive.

\subsection{Reducing the initial embryo mass}

\begin{figure}
\includegraphics[width=\columnwidth]{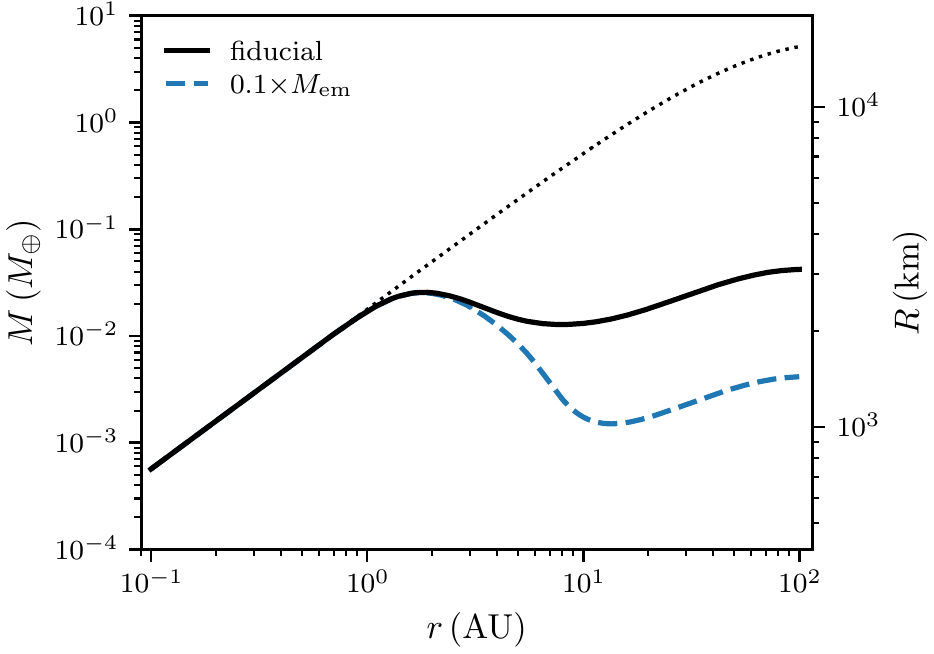}
\caption{Final sizes of embryos for reduced initial embryo mass. The embryo mass after $10$ Myr is shown as a function of distances for a $1\,M_\odot$ star for a $10$ times lower initial embryo mass (blue line). The fiducial case (black line) as well as the filament mass (dotted line) are shown for comparison.}
\label{fig:finalsizes_me01}
\end{figure}

In the fiducial run and variations thereof, we used the single most massive body from the streaming instability IMF as the embryo. In this case, the embryo is typically a factor $10^3$ more massive than the planetesimals (see Fig.~\ref{fig:initialsizes}). We run a model where we reduced the embryo mass by a factor of $10$ while keeping the total mass of the filament fixed. The mass ratio of embryo to planetesimals is hence ${\sim}10$ to $100$. Figure~\ref{fig:finalsizes_me01} shows that reducing the initial mass of the embryo does not change the final outcome for ${\lesssim}3\,\mathrm{AU}$, where the embryo accretes nearly all the available mass. Outside ${\sim}3\,\mathrm{AU}$, however, accretion efficiency decreases and for ${\gtrsim}20\,\mathrm{AU}$, the embryo does not grow significantly.

\subsection{Fragmentation, eccentricities, and diffusive widening}

\begin{figure}
\includegraphics[width=\columnwidth]{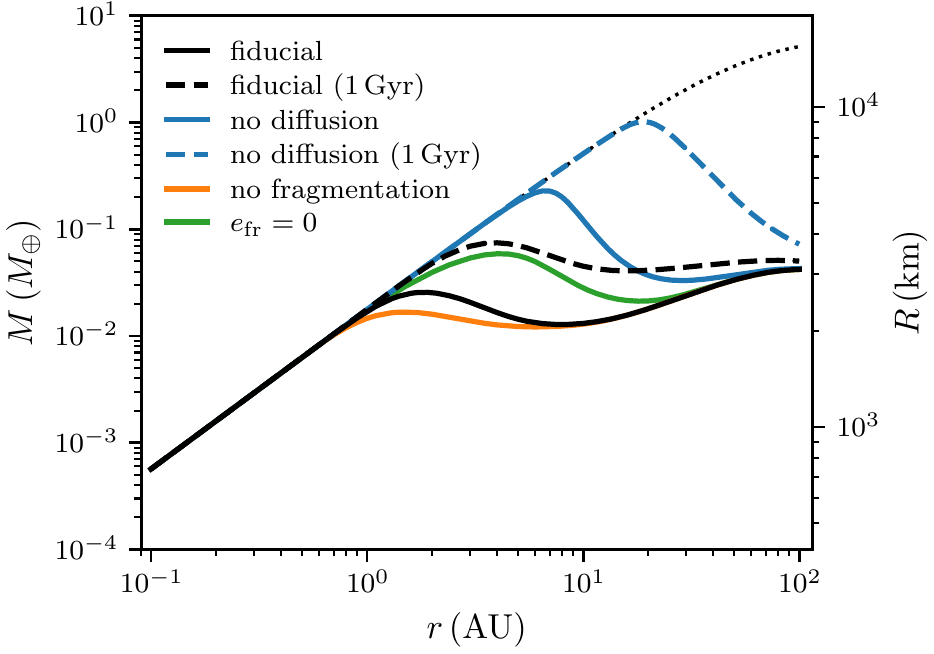}
\caption{Final sizes of embryos for different parameter variations. The embryo mass after $10$ Myr is shown as a function of distances for a $1\,M_\odot$ star for different cases (coloured lines) of no fragmentation, no diffusive widening, and fixed zero fragment eccentricity. The fiducial case (solid black line) as well as the filament mass (black dotted line) are shown for comparison.}
\label{fig:finalsizes_params}
\end{figure}

Figure~\ref{fig:finalsizes_params} compares the final masses of embryos for models, where we set the fragment eccentricity and inclination to zero, disabled fragmentation, disabled diffusive widening of the planetesimal and fragment rings, or extended the simulation from $10\,\mathrm{Myr}$ to $1\,\mathrm{Gyr}$.

Disabling fragmentation results in longer growth timescales. The final mass of the embryo after $10\,\mathrm{Myr}$, however, is not strongly affected. Inside of ${\sim}1\,\mathrm{AU}$, we find the same final mass while between ${\sim}1$ and ${\sim}5\,\mathrm{AU}$, the final mass is lower by less than a factor of $2$ at most. Outside ${\sim}5\,\mathrm{AU}$, we find the same final mass as in the fiducial case because fragmentation does not play a role.

The eccentricity of the fragments affect the growth behaviour more strongly. Fixing the fragments on orbits with zero eccentricity and inclination (that is assuming that gas drag is very efficient) allows the embryo to accrete fragments at a constant rate in the low-velocity regime \citep{Inaba2001,Chambers2006a}, while in the fiducial case the accretion rate decreases as fragments are excited by viscous stirring through the embryo and the planetesimals. As a consequence, we find that the embryos are more massive than in the fiducial case out to distances of ${\sim}20\,\mathrm{AU}$. We also run a model where we set the eccentricity and inclination of the embryo to zero. In this case, we did not find any significant deviation from the fiducial run. We conclude that a circular and planar embryo orbit as used in other studies \citep[e.g.][]{Chambers2006a} is a valid approximation because the eccentricity of the embryo is ${\sim}10^{-3}{\ll}e_\mathrm{p},e_\mathrm{fr}$ because of dynamical friction and gas drag (Fig.~\ref{fig:eccentricitymap}).

Lastly, we look at the diffusive widening of the planetesimal and fragment rings. Disabling diffusion has a strong impact on embryo growth. Within $10\,\mathrm{Myr}$ and out to ${\sim}5\,\mathrm{AU}$, the embryo accretes all the filament mass which allows growth to up to ${\sim}0.2\,M_\oplus$. On much longer timescales than $10\,\mathrm{Myr}$, embryos would grow up to ${\sim}1\,M_\oplus$ out to ${\sim}20\,\mathrm{AU}$. This is seen in Fig.~\ref{fig:finalsizes_params} where we show the final mass after $1\,\mathrm{Gyr}$ for the fiducial and the no-diffusion case for comparison. Without diffusion embryos grow up to ${\sim}1\,M_\oplus$, whereas with diffusion even after $1\,\mathrm{Gyr}$ the mass is at most ${\sim}0.1\,M_\oplus$. The reason for the strongly enhanced growth without diffusive widening is that the surface densities of planetesimals and fragments decrease only through accretion. However, the increase of eccentricities and inclinations through viscous stirring cause the bodies to occupy a larger volume additionally reducing the surface density and hence reducing the accretion rate of the embryo which is proportional to the surface density of the accreted bodies.


\subsection{Implications for pebble accretion}

The growth of embryos by planetesimal accretion in the filaments formed through streaming instability is efficient only in the inner part of the protoplanetary disc. In the inner disc, the collision timescale is short enough and fragmentation of planetesimals is efficient enough for an embryo to accrete all the available material. At larger distances, and especially outside ${\sim}5$ to $10\,\mathrm{AU}$, planetesimal accretion is highly inefficient, even though the available material in the filament increases. The larger sizes of the planetesimals, the excitation of planetesimal eccentricities, and the lack of fragmentation prevents embryos from growing massive within the lifetime of the disc of $10\,\mathrm{Myr}$.

The inefficient growth by planetesimal accretion does not necessarily imply that planets cannot form at all. We did not consider pebble accretion in our model because we focused on the accretion of the filament material turned into planetesimals. However embryos might still be able to reach masses for which pebble accretion becomes an highly efficient growth process. Pebble accretion becomes important when the friction time of the pebbles is shorter than the time in which they would pass by the embryo \citep{Ormel2017}. This condition leads to the onset mass for pebble accretion
\begin{align}
M_\mathrm{on}&=\frac{1}{4}\tau_\mathrm{s}\eta^3M_\star \nonumber \\
&\approx 4.871\times10^{-7}\,M_\oplus\,\left(\frac{\tau_\mathrm{s}}{0.01}\right)\left(\frac{M_\star}{M_\odot}\right)^{-17/7}\left(\frac{L_\star}{L_\odot}\right)^{6/7}\left(\frac{r}{\mathrm{AU}}\right)^{12/7}
\label{eq:onsetmass}
\end{align}
\citep{Visser2016,Ormel2017}. Above the so-called transition mass, pebble accretion becomes very efficient. The transition mass marks the change from drift-driven (Bondi) accretion to shear-driven (Hill) accretion of pebbles \citep{Lambrechts2012,Johansen2017,Ormel2017}. That means that in the latter case pebbles from the entire Hill sphere are accreted by the embryo. The transition mass can be found by equating the Bondi radius $r_\mathrm{B}{=}GM_\mathrm{em}/(\eta v_\mathrm{K})^2$ and the Hill radius and reads
\begin{align}
M_\mathrm{tr}&=\frac{1}{\sqrt{3}}\eta^3M_\star \nonumber \\
&\approx1.125{\times}10^{-4}\,M_\oplus\,\left(\frac{M_\star}{M_\odot}\right)^{-17/7}\left(\frac{L_\star}{L_\odot}\right)^{6/7}\left(\frac{r}{\mathrm{AU}}\right)^{12/7}
\label{eq:transitionmass}
\end{align}
\citep{Ormel2017}. Pebble accretion stops when the embryo reaches the pebble isolation mass. At this mass, the embryo carves a gap in the gas disc that creates a pressure bump outside its orbit which stops pebbles from drifting inward and being accreted by the embryo. The pebble isolation mass is
\begin{align}
M_\mathrm{iso}&=25\,M_\oplus \nonumber \\
&\times\left(\frac{H/r}{0.05}\right)^3\left(0.34\left(\frac{-3}{\log_{10}\alpha_\mathrm{t}}\right)^4+0.66\right)\left(1-\frac{\frac{\partial\ln P}{\partial\ln r}+2.5}{6}\right),
\label{eq:isolationmass}
\end{align}
which is derived from fits to hydrodynamic simulations of pebble accretion \citep{Bitsch2018}.

\begin{figure*}
\includegraphics[width=\textwidth]{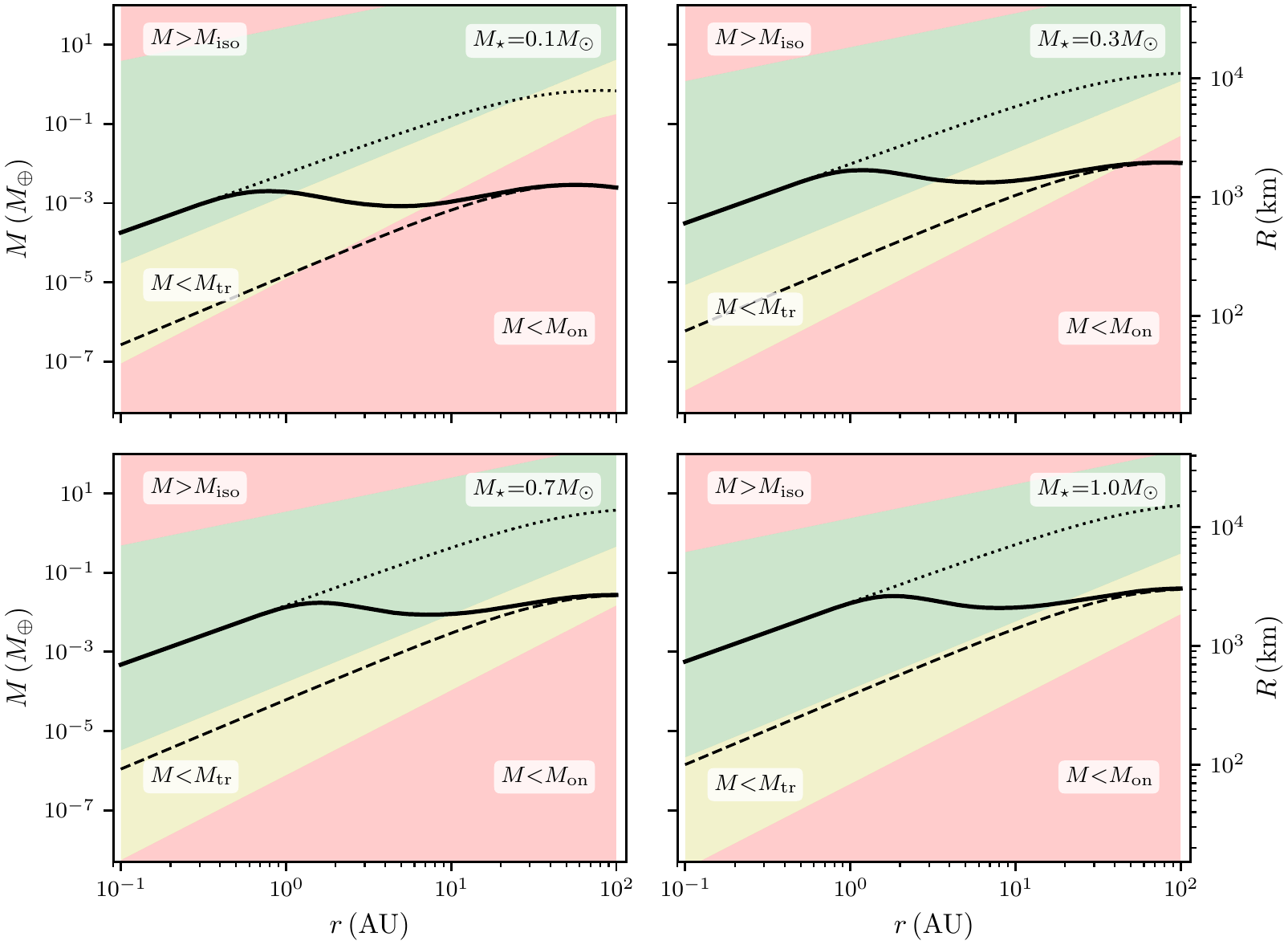}
\caption{Mapping final embryo masses to pebble accretion regimes. The figure shows the masses of the embryos at $10\,\mathrm{Myr}$ for four different stellar masses as a function of distance. The initial masses (dashed lines) and the filament masses (dotted lines) are shown for comparison. Where the mass is below the onset mass $M_\mathrm{on}$ (Eq.~\ref{eq:onsetmass}) or above the pebble isolation mass $M_\mathrm{iso}$ (Eq.~\ref{eq:isolationmass}), pebble accretion would be absent (red area). Above the onset mass but below the transition mass $M_\mathrm{tr}$ (Eq:~\ref{eq:transitionmass}), embryos would accrete pebbles on the Bondi branch (yellow area). Efficient pebble accretion would be possible for embryo masses above $M_\mathrm{tr}$ (green).}
\label{fig:allfinal_10Myr}
\end{figure*}

We now compare the embryo masses to the characteristic masses for pebble accretion given above. Figure~\ref{fig:allfinal_10Myr} shows a map where we highlight the different regimes of pebble accretion. For an embryo to accrete pebbles efficiently, the mass needs to be above the transition mass. Below the transition mass and above the onset mass, embryos would still be able to accrete pebbles, but on the less efficient Bondi branch. We see from Fig~\ref{fig:allfinal_10Myr} that the initial embryo mass is below the transition mass for all stellar masses; even though the difference is small for $M_\star{=}1\,M_\odot$. For $M_\star{=}0.1\,M_\odot$, the initial embryo mass is even below the onset mass for distances ${\gtrsim}2\,\mathrm{AU}$. For $M_\star{=}0.3\,M_\odot$, the initial embryo mass is below the onset mass outside of ${\sim}50\,\mathrm{AU}$. We also see from Fig~\ref{fig:allfinal_10Myr} that the maximum mass an embryo can reach through planetesimal accretion in a filament (the filament mass $M_\mathrm{fil}$) is above the transition mass (except for the $M_\star{=}0.1\,M_\odot$ outside ${\sim}20\,\mathrm{AU}$), but well below the pebble isolation mass, which means that there would be enough mass in planetesimals available for embryos to grow into the pebble accreting regime. However, we find that embryos growing through accretion of planetesimal would reach the transition mass only out to a distance of ${\sim}20\,\mathrm{AU}$ for a solar-like central star within the lifetime of a protoplanetary disc. For stars of lower mass, this distance shifts significantly inwards to ${\lesssim}1\,\mathrm{AU}$ for $M_\star{=}0.1\,M_\odot$. Therefore, we conclude that planetesimal accretion might be a channel for forming the seeds for pebble accretion out to ${\sim}20\,\mathrm{AU}$. Farther out, where planetesimal accretion becomes negligible, pebble accretion even though on the slow Bondi branch, would be the only growth channel. Our result is in agreement with \citet{Liu2019} who investigated the growth of planetesimals to planets at a single site, namely the water snowline at $2.7\,\mathrm{AU}$, through planetesimal and pebble accretion using $N$-body simulations. Also in their work, embryos would grow to masses of $10^{-3}$ to $10^{-2}\,M_\oplus$ through planetesimal accretion after which pebble accretion would take over. Comparing the final embryo masses in our model at $2.7\,\mathrm{AU}$ (\ref{fig:massmap} and the bottom right panel of \ref{fig:allfinal_10Myr}), shows comparable masses.

\subsection{Limitations of the model}

Our model describes the growth of an embryo at a fixed location. We neglected the migration of the embryo, the planetesimals, and the fragments for several reasons:
\begin{enumerate*}[label=(\roman*)]
\item for the embryo masses considered here (${\lesssim}0.1\,M_\oplus$), the migration timescales are ${\sim}\mathrm{Myr}$ and longer \citep{Tanaka2002,Cresswell2008,Ida2020},
\item apart from having a more complicated model, radial drift of planetesimals and fragments would only reduce the accretion efficiency due to an additional drain of available material, and
\item gas drag induced radial drift peaks for metre-sized bodies but planetesimals, fragments, and embryos have sizes of kilometre to several hundreds of kilometres resulting in slow radial drift.
\end{enumerate*} 
Therefore, having non-migrating bodies provides us an upper limit on the final masses, any migration would result in bodies with lower mass.

In our study, we did not assume that streaming instability forms filaments at special locations in the disc, such as snowlines where a pressure bump would naturally lead to an increased solid-to-gas ratio due to a pile-up of pebbles that would trigger the streaming instability \citep{Drazkowska2017,Schoonenberg2018}. Instead, we look at what would happen if filaments occur at any location \citep{Carrera2017,Lenz2019}. A consequence of this would be a significant reservoir of planetesimals that are not accreted. This reservoir can nevertheless interact with the planets that might form by pebble accretion leading to scattering and populating of the Kuiper belt, scattered disc, and Oort cloud, thus providing the bodies for comets and Kuiper belt objects \citep{Brasser2013}. The fact that outside of $10\,\mathrm{AU}$ neither significant growth nor fragmentation occurs might imply that also the size distribution of the planetesimals remains largely unchanged. The cold-classical Kuiper belt might be a remnant of this. In our model, the total mass of planetesimals in the whole disc is ${\sim}333\,M_\oplus$. The mass contained in the asteroid belt region between $2\,\mathrm{AU}$ and $4\,\mathrm{AU}$ is ${\sim}10\,M_\oplus$, and in the region of the primordial disc region between $15\,\mathrm{AU}$ and $30\,\mathrm{AU}$ it is ${\sim}50\,M_\oplus$, both of which are orders of magnitude larger than the current mass in asteroids and in the Kuiper belt, estimated to be $4{\times}10^{-4}\,M_\oplus$ and ${\sim}10^{-2}\,M_\oplus$, respectively \citep{DeMeo2013,Fraser2014}. Therefore, efficient depletion becomes necessary, such as the giant planet instability in the Nice model, which could have been responsible for sculpturing the outer Solar System and depleting the asteroid belt by scattering of planetesimals and ejection of planetesimals from the Solar System \citep{Gomes2005,Tsiganis2005,Morbidelli2010,Brasser2013}. On the other hand, we assumed that $100\,\%$ of the filament mass is converted to planetesimals, which gives an upper limit on the available mass. The planetesimal formation efficiencies in streaming instability simulations are not well constrained and can vary significantly from ${\lesssim}10\,\%$ to as high as ${\sim}80\,\%$ \citep{Abod2019}. Converting less pebbles to planetesimals will reduce the available mass significantly and, additionally, lifting the assumption that filaments form throughout the disc reduces the amount of planetesimals even further.

We furthermore neglect that filaments might interact with each other and that, especially the large embryos in the outer disc that might have Hill radii exceeding the typical spacing of the filaments, would be able to accrete from neighbouring filaments. However, the relevance of this might be low because even though there is a huge mass reservoir of planetesimals of several Earth masses, the embryos accrete almost no planetesimals.


\section{Conclusion}
\label{sec:conclusions}

In this paper, we modelled the planetesimal accretion phase that follows the birth of planetesimals. Therefore, we developed a model for the growth of a large planetesimal (embryo) embedded in a population of smaller planetesimals of characteristic size. The model included mass growth of the embryo, the fragmentation of planetesimals and the velocity evolution of all involved bodies in a self-consistent fashion. We represented the planetesimal size distribution at birth with bodies of characteristic masses, the planetesimals and the embryo. Our growth model hence described an oligarchic-like growth. Fragmentation assumed a representative fragment size. We found that embryos accrete the available material efficiently only in the inner disc where a combination of high planetesimal surface density and fragmentation ensures short growth timescales for the embryo. On the other hand, we find little to no growth in the outer parts of the disc beyond ${\sim}5$ to $10\,\mathrm{AU}$ on a $10\,\mathrm{Myr}$ timescale. The embryos typically reached masses in the range ${\sim}10^{-3}$ to $10^{-1}\,M_\oplus$. When we compare the embryo masses to the transition mass for pebble accretion, we find that embryos would be able to grow into the pebble accreting regime through planetesimal accretion out to ${\sim}20\,\mathrm{AU}$. Pebble accretion on the less efficient Bondi branch might help embryos to reach the transition mass also beyond ${\sim}20\,\mathrm{AU}$.

\begin{acknowledgements}
We thank the anonymous referee for a constructive feedback that contributed in improving the quality of our work. A.J. is supported by the Swedish Research Council (Project Grant 2018-04867), the Danish National Research Foundation (DNRF Chair grant DNRF159), and the Knut and
Alice Wallenberg Foundation (Wallenberg Academy Fellow Grant 2017.0287).
A.J. further thanks the European Research Council (ERC Consolidator Grant 724
687-PLANETESYS), the Göran Gustafsson Foundation for Research in Natural Sciences and Medicine, and the Wallenberg Foundation (Wallenberg Scholar
KAW 2019.0442) for research support.
\end{acknowledgements}

\bibliographystyle{bibtex/aa} 
\bibliography{../references/ref} 

\begin{appendix}
\end{appendix}

\end{document}